\documentclass{PoS}

\usepackage{amsmath}

\title{First results in QCD with 2+1 light flavors using the fixed-point
  action} 

\ShortTitle{First results in QCD with 2+1 light flavors using the fixed-point
  action} 

\author{Anna Hasenfratz\\
  Department of Physics, University of Boulder, Boulder, CO-80304-0390, USA}

\author{Peter Hasenfratz, Ferenc Niedermayer\\
  Institute for Theoretical Physics, University of Bern, CH-3012 Bern,
  Switzerland} 

\author{\speaker{Dieter Hierl} and Andreas Sch\"afer\\
  Institute for Theoretical Physics, University of Regensburg, D-93040
  Regensburg, Germany\\
  E-mail: \email{dieter.hierl@physik.uni-regensburg.de}}

\abstract{This is a progress report on 2+1 flavor simulation with the FP
  action on a lattice with spatial size $L=1.2\,\mathrm{fm}$. Since $m_{ud}$ 
  is quite small in our simulation we are in the delta regime for the two
  light flavors where the low lying excitations are described by a quantum
  mechanical rotator. From here we extract the low energy constant $F$.
  We also measure the AWI mass and present results  on numerical issues like low-mode averaging and
  autocorrelation times.}

\FullConference{XXIVth International Symposium on Lattice Field Theory\\
  July 23-28, 2006\\
  Tucson, Arizona, USA}

\begin{document}

\section{Introduction}
We performed simulations for $2+1$ light flavors Lattice QCD with the
fixed-point Dirac operator.
The (parametrized) fixed-point operator we use approximatively satisfies the
Ginsparg-Wilson relation \cite{Ginsparg:1981bj}:
\begin{equation}
  D \gamma_5 + \gamma_5 D = D \gamma_5 2 R D\;,
\end{equation}
and is limited to a hypercube \cite{Hasenfratz:1997ft}.
(Here $R$ is a non-trivial local matrix.) 
The quark mass is introduced by
\begin{equation}
  D(m)=D+m\left(\frac{1}{2R}-\frac{1}{2}D\right).
\end{equation}
The fixed-point action gave very promising results in the quenched
approximation, in particular good scaling even at $a = 0.15\,\mathrm{fm}$ 
\cite{Gattringer:2003qx,Hasenfratz:2002rp}.
We tuned the coupling in our full QCD simulation to be close to 
$a = 0.15\,\mathrm{fm}$ \cite{Hasenfratz:2005tt}, and measurement of the Sommer parameter $r_0$ gave a
result in agreement with this value.
We simulate two lattices, of size $8^3\times24$ and $12^3\times24$ at this
coupling. The results presented here refer to the smaller lattice.

\section{Autocorrelation times}

\begin{figure}[b]
  \centerline{
    \includegraphics[width=.6\textwidth]{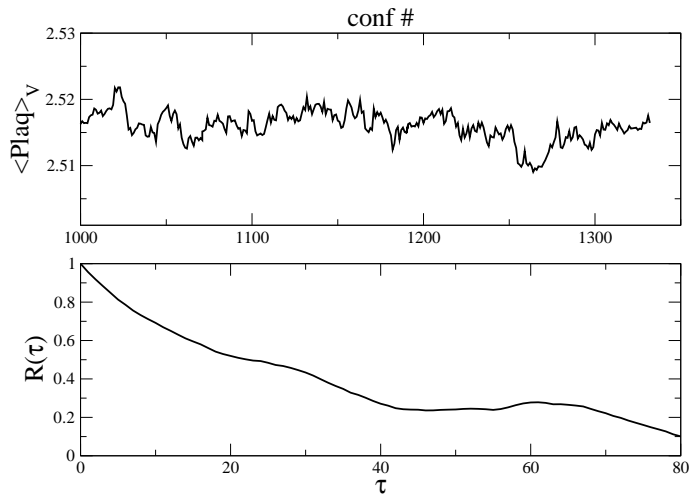}
  }
  \caption{{}MC history for the smeared plaquette and the corresponding
    autocorrelation function.} 
  \label{fig1}
\end{figure}

We estimated the autocorrelation time of plaquettes built from smeared links
and that of the pseudoscalar-pseudoscalar correlator. We are using a global
update \cite{Hasenfratz:2005tt}, the configurations are separated by 
$\approx 0.7$ standard Metropolis sweeps.

Fig.~\ref{fig1} shows the MC history for the plaquette  for a longer run and
the autocorrelation function extracted from it. This gives an autocorrelation
time $\tau_0 \approx 40$.

\begin{figure}[t]
  \centerline{
    \includegraphics[width=.6\textwidth]{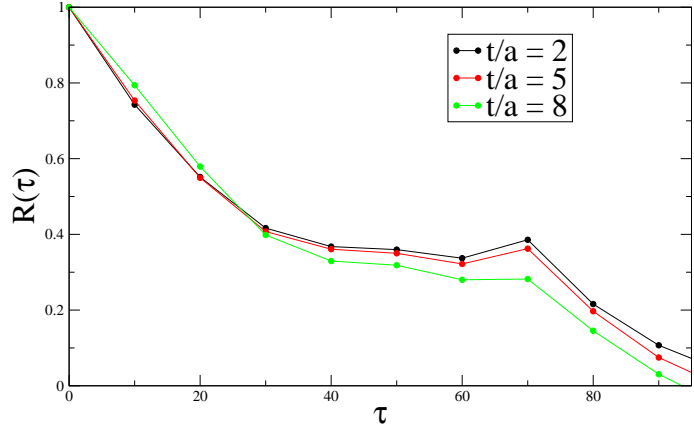}
  }
  \caption{{}The autocorrelation function for the $\langle P(0)P(t)\rangle$
    correlator at different $t$ values. }
  \label{fig2}
\end{figure}

In Fig.~\ref{fig2} we plot the autocorrelation function for the zero-momentum 
$\langle P(0)P(t)\rangle$  correlator at different time separations $t$.
The autocorrelation time for this quantity is estimated to be somewhat higher.

\newpage

\section{The delta regime}

The epsilon regime is defined by:
\begin{equation}
  m_{\mathrm{PS}} L,\;m_{\mathrm{PS}} L_4  \lesssim 1\,,
\end{equation}
and the box size should be large enough, $2\pi F L \gg 1$, for 
the validity of chiral perturbation theory.
Here $m_{\mathrm{PS}}$ is the pseudoscalar mass in infinite volume.

In the delta regime \cite{Leutwyler:1987ak} one considers a lattice under
similar conditions but elongated in the time direction so that the low-lying
spectrum in the corresponding spatial volume can be measured.
In this regime, for two massless quarks, the low lying spectrum is given by an
O(4) rotator having in leading order in $1/L$ a mass gap
\begin{equation} \label{meff}
  m_\mathrm{eff}= \frac{3}{2F^2 L^3}
\end{equation} 
Note, however, that for our box size of $L=1.2\,\mathrm{fm}$ one expects
non-negligible corrections to this leading order result \cite{Hasenfratz:1993vf}.

\section{Low-mode averaging}

We used the method of low-mode averaging \cite{Edwards:2001ei,DeGrand:2004qw, Giusti:2004yp}
to decrease the error in the zero-momentum meson correlators
\begin{equation}\label{meson}
  C(t;y) = \left.\left\langle \sum_{\vec{x}} \mathrm{tr}
      \left( \Gamma_1 \gamma_5 G(x,y)^\dagger \gamma_5 \Gamma_2 G(x,y) \right)
    \right\rangle \right|_{t=x_4-y_4} \,,
\end{equation}
where $\Gamma_1$, $\Gamma_2$ are arbitrary Clifford algebra matrices. As part of our updating algorithm we calculate the low-lying eigenvalues 
and their corresponding eigenvectors \cite{Hasenfratz:2005tt} which
allows us to perform such averages. 
For the $8^3 \times 24$ lattice we stored  $48$
eigenvectors with the smallest eigenvalues. We split up the correlator in 
Eq.~(\ref{meson}) into three parts:
\begin{equation}
  C(t;y) = C_{ll}(t) + C_{lh}(t;y_0) + C_{hh}(t;y) \,,
\end{equation}
where $l$ refers to the low-mode part $G_l(x,y)$ of the quark propagator
(which is given by the calculated eigenvalues and eigenvectors), and $h$ to
the rest of the propagator, $G(x,y) = G_l(x,y) +  G_h(x,y)$.
We analyze the effects of low-mode averaging for $C_{ll}$ and $C_{lh}$
separately.
\begin{figure}
  \centerline{
    \includegraphics[width=.7\textwidth]{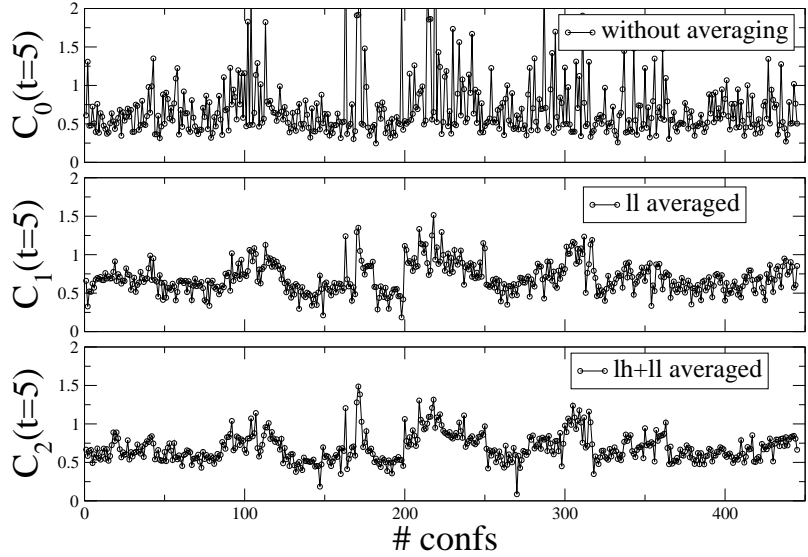}
  }
  \caption{{}The zero-momentum $\langle P(0)P(t=5) \rangle$ correlator
    measured at fixed source position ($C_0$);
    by using the eigenvectors to average the source position  over the whole
    lattice in the low-low part  ($C_1$); by averaging in the source position
    over a time-slice in the low-high part as well ($C_2$).}
  \label{fig3}
\end{figure}

On  Fig.\ref{fig3} we compare the zero momentum correlator $\langle
P(0)P(t)\rangle$  at time separation $t=5$ for the different amounts of
averaging. $C_0$ is the correlator without any averaging over the source
position, $C_1$ is when the $ll$ part is averaged over the whole volume, $C_2$
is when in addition the $lh$ part is also averaged over a time slice.
While the $ll$ averaging is very effective, the $lh$ averaging barely shows
any suppression of the fluctuations. Since the latter one is also more
expensive, one can safely ignore this option and average only in the $ll$ part.

\section{Results}

\subsection{The PCAC mass}

\begin{figure}
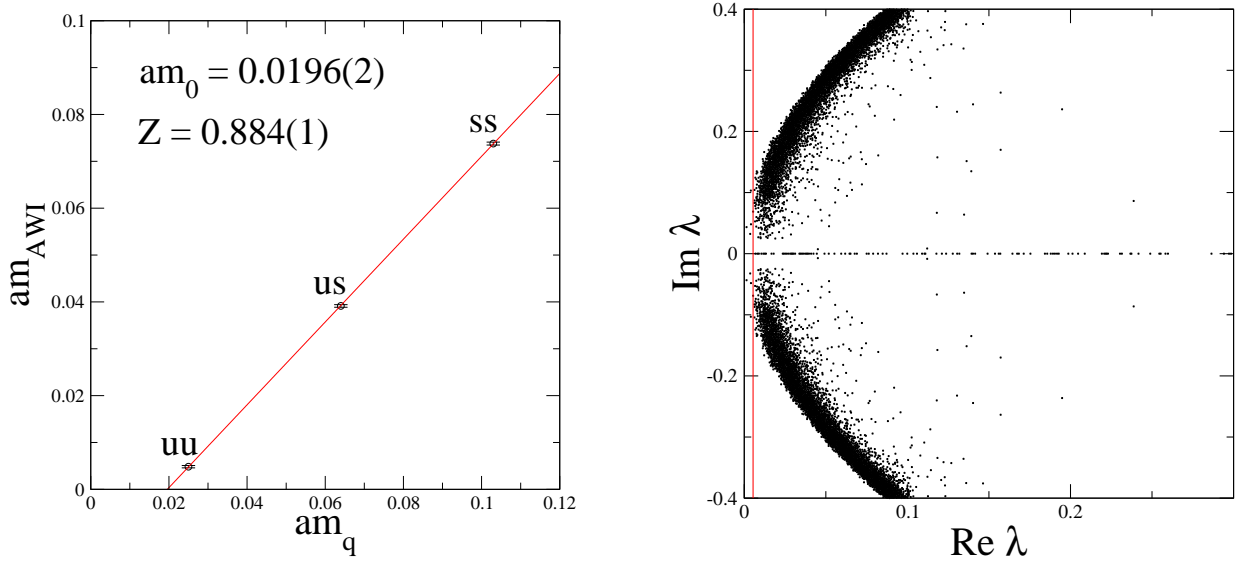

  \centerline{
    \includegraphics[width=.5\textwidth]{awi_mass_chi_x.eps}
    \hspace{10mm}
    \includegraphics[width=.5\textwidth]{ev_all_x.eps}
  }
  \caption{{\bf Left:} The chiral extrapolation of $m^\mathrm{AWI}(m_q)$ using
    the combinations ud, us, and ss for the quark flavors. We find an additive
    mass renormalization of $am_0=0.0196(2)$ and a renormalization constant of
    $Z=0.8837(7)$. {\bf Right:} The low-lying eigenvalues of the massive Dirac
    operator for $am_u=0.025$. The red line shows the quark mass with this
    mass shift subtracted, $am_{\mathrm{sub}}=a m_q - a m_0$.}
  \label{fig6}
\end{figure}

Because we are using the parametrized fixed-point action which 
solves the Ginsparg-Wilson-equation only approximately 
we get an additive mass renormalization for our PCAC masses.
The size of this mass shift is an indicator for the quality of our
approximation. From preliminary runs \cite{Hasenfratz:2005tt} we estimated this additive
mass renormalization to be around $0.02$. We ran the present simulations with
lattice quark masses $am_{ud}=0.025$, $am_s=0.103$.
To determine the actual additive mass renormalization 
we calculated ratios
\begin{equation}\label{ratio}
  \rho_{ab}(t) = \frac{\langle0|\partial_4^*A_4^{ab}(t)P_{ba}(0)|0\rangle}
  {2\langle0|P_{ab}(t)P_{ba}(0)|0\rangle}\,, \qquad a,b=u,d,s \,.
\end{equation}
For both correlator functions we used low-mode averaging and found
well-defined plateaus, typically beyond $t\approx 5$. 
These we denote by $m^\mathrm{AWI}_{ab}$. Plotting $m^\mathrm{AWI}_{ab}$ against $m_q \equiv (m_a+m_b)/2$ and
extrapolating in $m_q$  to $m^\mathrm{AWI}=0$ we obtain the mass shift $m_0$.
Using naive currents, i.e. the non-conserved axial-vector current in the enumerator and the
pseudoscalar density in the denominator of Eq.~\eqref{ratio}, we also get a multiplicative renormalization factor $Z$:
\begin{equation}
  m^\mathrm{AWI}_{ab} = Z(m_q - m_0)\,, \qquad m_q = \frac{1}{2}(m_a+m_b) \,.
\end{equation}
As seen on the left hand side of Fig.~\ref{fig6} the actual values
are described very well by this linear dependence. We also plotted on the right hand side of Fig.~\ref{fig6} the $48$ low-lying eigenvalues we stored for several of our configurations with $am_u=0.025$ on
our  $8^3\times24$ lattice.
We find that they are lying approximatively on a Ginsparg-Wilson circle
shifted by the subtracted mass 
$am_{\mathrm{sub}} = am_u-am_0= 0.0054(2)$ 
(corresponding to $7.2(4)\mathrm{MeV}$) away from 
$\mathrm{Re}\lambda=0$ (red line).
The subtracted lattice mass for the strange quark is
$am_s-am_0=0.083$, corresponding to $\approx 110\,\mathrm{MeV}$.

\subsection{Effective masses}

\begin{figure}
  \centerline{\includegraphics[width=.6\textwidth]{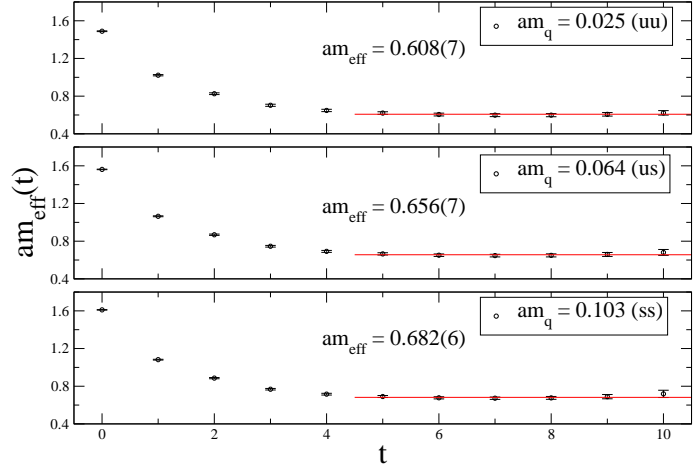}}
  \caption{{}Mass plateau in the $\langle P_{ab}(0)P_{ba}(t) \rangle$
    correlator, for $ab=ud, us, ss$.}
  \label{fig8}
\end{figure}

For the point-like pseudoscalar density $P=\overline{\psi}\gamma_5\psi$ we
calculated the zero-momentum $\langle PP \rangle$ correlator for three
different quark flavor combinations (uu,us,ss) and found very stable effective
mass plateaus shown in Fig.~\ref{fig8}.

\begin{figure}
  \centerline{
    \includegraphics[width=.6\textwidth]{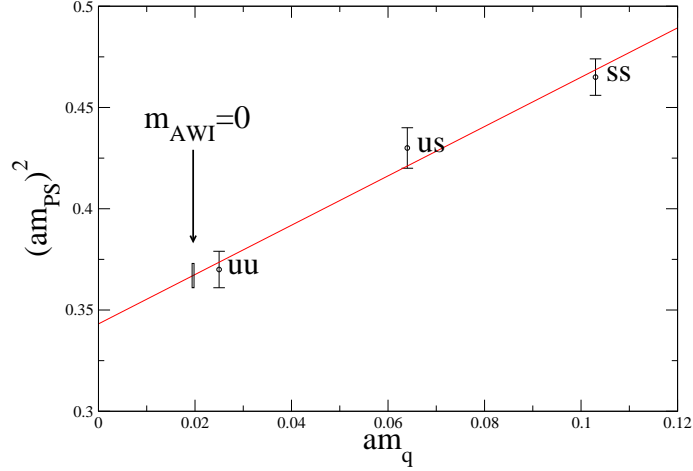}
  }
  \caption{{}The chiral extrapolations in the pseudoscalar channel. 
    On the right hand side we extrapolate $(am_{\mathrm{eff}})^2$ linearly in
    $am_q \to am_0$  and on the left hand side we do the same for
    $am_{\mathrm{eff}}$.} 
  \label{fig10}
\end{figure}

Linear chiral extrapolation of $m_\mathrm{eff}^2$ in $m_q$ to the chiral
limit, shown in Fig.~\ref{fig10}, gives:
\begin{equation}
  (am_\mathrm{eff})^2(m_0) = 0.367(6)\;.
  \label{eq5_3}
\end{equation}
Using the finite volume mass gap in lowest order of chiral perturbation
theory, Eq.~\eqref{meff}, this leads to $F=92.7(4)\,\mathrm{MeV}$
which agrees with the physical value of the pion decay constant, 
$F_\pi = 92.4\,\mathrm{MeV}$.
However, this agreement is presumably accidental, since the spatial extend of our
small lattices is uncomfortably small, and sizeable corrections to the leading
order result of Eq.~\eqref{meff} are expected.

\section{Summary and Outlook}

We calculated the autocorrelation for the $8^3 \times 24$ lattice and found 
autocorrelation times $\tau_0 \approx 40$ for plaquettes on smeared links and 
$\tau_0 \approx 80$ for the $\langle PP \rangle$ correlation functions at
fixed time distances.

In the measurement of the meson correlators the 
low-mode averaging was found to improve the signal significantly.
Since the low lying eigenmodes of the Dirac operator were available on each
configuration from our updating, this did not require extensive computing
resources.

We determined the additive mass renormalization and 
verified that our light quark masses are near to the chiral limit.
We found that the AWI mass depends nearly linearly on the average
of the corresponding lattice quark masses.
We found that the eigenvalues lie in good approximation on the shifted 
Ginsparg-Wilson circle. 

For the pseudoscalar spectrum we obtained the finite volume mass gap
in the chiral limit. Comparing it to the lowest order theoretical prediction
we obtained a value for the low energy constant $F$.
The surprising agreement with the physical value is presumably an
accident since finite volume effects could be quite large
at this lattice size.

In the near future we will also have access to larger lattices, on which we
plan to analyze conserved currents \cite{Hasenfratz:2002rp} and all-to-all
propagators \cite{Foley:2005ac,Burch:2006mb}.

\section{Acknowledgement}
This work was supported in part by Schweizerischer
Nationalfonds and by DFG (FG-465). We acknowledge the support and computing 
resources at CSCS, Manno and LRZ (project h032z).

\end{document}